\documentclass[a4paper,11pt]{article}
\usepackage{jheppub} 
\usepackage{lineno}
\nolinenumbers
\newcommand{\rmd}{\textrm{d}}

\usepackage{xcolor}
\usepackage{subcaption}
\usepackage{amsmath}
\usepackage{amssymb}
\usepackage{graphicx}

\usepackage[normalem]{ulem}

\title{\boldmath Frame-dependency of the confinement temperature in a strongly-coupled plasma under rotation: a holographic description}

\author{Nelson R. F. Braga}
\author{and Alexsandre L. Ferreira Jr.}
\affiliation{UFRJ — Universidade Federal do Rio de Janeiro,\\ Instituto de Física, Caixa Postal 68528, Rio de Janeiro, Brasil}

\emailAdd{braga@if.ufrj.br}
\emailAdd{alexsandrej@if.ufrj.br}

\abstract{Recently, it was demonstrated that the disagreement between lattice calculations and holographic models, regarding rotational effects in the quark-gluon plasma (QGP), occurs due to different choices of reference frames. While a static observer measures a confinement temperature that decreases with rotation, one in a co-rotating frame finds the opposite behavior. Both results are correct, and a comprehension of the frame-dependency of the critical temperature is a significant step in the description of the QGP under rotation. In this article, we generalize this previous holographic result by describing the plasma through the most general Myers-Perry black hole solution. This breaks the spherical symmetry present in the equal angular momentum case, considered before. As a consequence, the local temperature measured by a co-rotating observer has a non-trivial angular dependence: it can decrease, increase or even behave non-monotonically when rotational velocity increases.}

\begin{document}
\maketitle
\flushbottom

\section{Introduction}
\label{intro}

Laboratory production of the quark-gluon plasma, at the Relativistic Heavy Ion Collider (RHIC) and at the Large Hadron Collider (LHC), inaugurated an important window of observation in the strongly-coupled regime of quantum chromodynamics (QCD), which has been widely scrutinized, resulting in several important break troughs over the last decades \cite{Shuryak2008,Casalderrey-Solana:2011dxg,Busza:2018rrf}. Particularly, a good amount of effort has been applied to reconstruct the phase diagram of QCD under the conditions produced in heavy ion collisions.  Such as large magnetic fields \cite{Ballon-Bayona:2013cta,Mamo:2015dea,Dudal:2015wfn,Rodrigues:2017cha,Braga:2019yeh,Braga:2020hhs,Bohra:2020qom} and, more recently, angular momenta \cite{CZ21,BFJ22,Y23,WF24,ZH23,FFH21,CFS22,CFS24,FS25,WCHR25,BKKR21,BKKR22,CGM23,MT23,SSWXH24,CSDH25,CCDH25,GHZ23,J24}.

Non central heavy ion collisions generate plasmas spinning with relativistic angular velocities of the order of $\Omega\sim 7\, \textrm{MeV}$ \cite{STAR2017}, putting a spotlight on the effects of rotation in strongly-coupled systems, especially in the last years. Nonetheless, despite considerable efforts, the effects of angular momentum in the confinement/deconfinement transition are yet poorly understood. More than that, we arrived at a conundrum, as different approaches disagree on the qualitative behavior of the confinement critical temperature. While some effective models (including previous holographic descriptions \cite{CZ21,BFJ22,Y23,WF24,ZH23}), and perturbative calculations report a decrease in the critical temperature with rotation \cite{FFH21,CFS22,CFS24,FS25,WCHR25}; others, including lattice QCD (LQCD) results, show the opposite behavior \cite{BKKR21,BKKR22,CGM23,MT23,SSWXH24,CSDH25,CCDH25}; alongside, even non-monotonic profiles were found \cite{GHZ23,J24}. Currently, intense research in the field has been done to overcome this riddle and comprehend the physics of confinement in spinning strongly-coupled systems. 

There are different recipes to introduce angular moment. In lattice calculations, rotation is introduced in the spacetime metric by treating non-inertial rotational forces as gravitational fields, a clever use of the equivalence principle \cite{BKKR21,BKKR22,CGM23}. This implies that calculations are being performed in the reference frame co-rotating with the plasma, in contrast to a static one. On the other hand, previous holographic models include rotation through a boost in a compact coordinate, i.e., considering solely inertial effects, while calculations are performed in a static reference frame \cite{CZ21,BFJ22,Y23,WF24,ZH23}. 

Remarkably, these different choices of reference frames might be responsible for the disagreement between LQCD and some holographic models, as reported in Ref. \cite{Braga:2025wox}, a result that gives a meaningful step towards reconciliation. There, a holographic model is presented in which the confinement temperature increases with angular velocity when measured by an observer co-rotating with the plasma, in qualitative agreement with LQCD; while it decreases when calculated in a static frame. 

The difference in temperatures observed by reference frames with relative accelerated motion is not a novelty. This is a direct consequence of the covariant formulation of the Fick's law for heat diffusion \cite{SV19}, and it was demonstrated almost a century ago by Tolman and Ehrenfest \cite{T30,TE30}. Therefore, it is noteworthy that such well known relativistic effect lies in the root of an apparent contradiction between previous holographic models and lattice results. Nonetheless, despite different temperatures measurements, it must be noted that the system is either in a plasma or in a hadronic state, independent of observer.
The holographic results mentioned above were obtained through the anti-de Sitter/conformal field theory (AdS/CFT) correspondence. Holography became a valuable tool to investigate strongly-coupled systems in $D$ dimensions, such as the QGP, by relating their observables with the ones calculated in a weakly-coupled gravitational system in $D+1$ dimensions \cite{GKP98,M99,GK09,W98,W98b}. As a consequence, it has been successfully applied to investigate several aspects of QCD and the QGP, for example in \cite{PS02,BB03,BB04,KSS05,GN08,GNPR08,DGR11,CNNPRR17,RGHNNPR24,HGMNNPRR24}.  In the version used here, a five dimensional asymptotic AdS spacetime corresponds to a $\mathcal{N}=4$ Super Yang-Mills (SYM) field theory in four dimensions. At finite temperatures, the super-symmetry is broken and the gauge theory displays several phenomena of interest present in QCD, including confinement \cite{W98}.

In special, the AdS/CFT correspondence allows for a geometrical description of the confinement/deconfinement phase transition through the Hawking-Page (HP) approach \cite{W98}, describing the change between an asymptotic AdS black hole (BH) and a thermal AdS spacetime \cite{HP83}. A resource fairly used to investigate the QCD phase diagram \cite{H07,BBNZ08}. Hence, with the purpose of accounting for non-inertial effects in the confinement temperature, Ref. \cite{Braga:2025wox} investigates the Hawking-Page phase transition in the five-dimensional equal angular moment Myers-Perry BH (MP BH)--a full spinning asymptotic AdS solution of Einstein's field equations. There, the dual theory corresponds to a rotating gauge theory in four dimensions with compact space-like hypersurfaces, usually given by three-spheres. The usual procedure to have a flat Minkowski boundary is to work with a black brane. However, the black brane regime, in which the BH radius is taken to be very large, i.e., $r_+\to\infty$, is not suited to investigate rotation, as any non-inertial effects are washed out \cite{ACKW24,Braga:2025wox}. 

The MP BH solution has two independent conserved angular momenta, given by the rotation parameters $(a,b)$, and it finds important applications in relativistic rotating hydrodynamics through the gauge/gravity duality (see the review \cite{ACKW24} and references therein). In the equal angular momenta case, where $a=b$, the spinning BH solution possess spherical symmetry, which simplifies most calculations. Nonetheless, it naturally raises the question: Which features of the holographic model would still be presented in the broken symmetry scenario? The main purpose of the present work is to answer this question. For that, we shall investigate the confinement/deconfinement transition through the HP approach in the most general MP BH solution, where the rotational parameters $(a,b)$ need not to be the same. 

Besides the importance of generalizing previous results, there are two main reasons for such task. First, in the equal angular momenta BH, the dual plasma is such that every point in the three-sphere has the same linear velocity, due to the spherical symmetry presented in the solution. This is in contrast to what is expected in the QGP produced by off-central collisions. 

Secondly, an alternative to the black brane limit in describing a rotating plasma in a flat spacetime is a stereographic projection from $R\times S^3$ to $R^4$. Considering the MP BH, a projection in the conformal boundary results in a flow with various features expected in the QGP produced by non-central collisions, denoted as Bantilan-Ishii-Romatschke flow \cite{Bantilan:2018vjv}. There, each rotation parameter inherited from the MP BH will represent a different quantity in the plasma's flux, therefore, it is important to control and understand each parameter separately.

In the next section the main features of the Myers-Perry solution are revisited, alongside with a suitable choice of coordinates to what follows. Moreover, the thermodynamic properties of the black hole are discussed. Then, Sec. \ref{sec3} presents the Hawking-Page phase transition from the MP BH to a pure thermal AdS space and its relation with the confinement/deconfinement transition. For that, the \textit{on-shell} action is calculated using a new technique for the computation of an integral, which might be useful in related works. The main discussion of the article is presented in Sec. \ref{sec4}, where we explain the concept of local temperature, and demonstrate that the confinement temperature calculated in the previous section amounts to the temperature measured in a static frame. Further, the co-rotating temperature, the one observed in a co-rotating frame, and is value in the confinement transition are introduced and discussed. Interestingly, due to the absence of spherical symmetry, one finds an angular dependence that is not observed in Ref. \cite{Braga:2025wox}. Finally, the article ends with some concluding remarks.

\section{Myers-Perry black hole}
\label{sec2}

In this work, rotation in a strongly-coupled plasma is investigated through a holographic description. For that, the gravitational dual to the plasma is chosen to be the Myers-Perry black hole, a five dimensional spinning uncharged black hole solution of general relativity with asymptotic AdS symmetry \cite{HHT99,Myers:1986un}. In this section we review the MP BH, discuss its main kinematical properties, with an emphasis in a suitable choice of coordinate system, and the thermodynamics of the solution. Further, we also comment on the black brane and equal angular momentum limits.

\subsection{Kinematical properties and coordinate systems}

In Boyer-Lindquist coordinates, the MP BH line element is given by:
 \begin{multline}
     \rmd s^2=\frac{(1+r^2/l^2)}{\rho^2r^2}\bigg(a\,b\,\rmd t-\frac{b(a^2+r^2)\mathrm{sin}^2(\tilde{\theta})}{\Xi_a}\rmd\tilde{\phi}-\frac{a(b^2+r^2)\mathrm{cos}^2(\tilde{\theta})}{\Xi_b}\rmd\tilde{\psi}\bigg)^2\\
     -\frac{\Delta_r}{\rho^2}\bigg(\rmd t-\frac{a\,\mathrm{sin}^2(\tilde{\theta})}{\Xi_a}\rmd \tilde{\phi}-\frac{b\,\mathrm{cos}^2(\tilde{\theta})}{\Xi_b}\rmd \tilde{\psi}\bigg)^2+\frac{\rho^2}{\Delta_{\tilde{\theta}}}\rmd \tilde{\theta}^2+\frac{\rho^2}{\Delta_r}\rmd r^2\\+\frac{\Delta_{\tilde{\theta}}\,\mathrm{sin}^2(\tilde{\theta})}{\rho^2}\bigg(a\,\rmd t-\frac{a^2+r^2}{\Xi_a}\rmd \tilde{\phi}\bigg)^2+\frac{\Delta_{\tilde{\theta}}\,\mathrm{cos}^2(\tilde{\theta})}{\rho^2}\bigg(b\,\rmd t-\frac{b^2+r^2}{\Xi_b}\rmd \tilde{\psi}\bigg)^2,
     \label{Bl_MP_ds2}
 \end{multline}
where
\begin{align}
    \Delta_r&=\frac{1}{r^2}(r^2+a^2)(r^2+b^2)\left(\frac{r^2}{l^2}+1\right)-2M,\\
    \Delta_{\tilde{\theta}}&=1-\frac{a^2}{l^2}\mathrm{cos}^2(\tilde{\theta})-\frac{b^2}{l^2}\mathrm{sin}^2(\tilde{\theta}),\\
    \rho^2&=r^2+a^2\mathrm{cos}^2(\tilde{\theta})+b^2\mathrm{sin}^2(\tilde{\theta})\\
    \Xi_a&=1-\frac{a^2}{l^2}, \qquad\Xi_b=1-\frac{b^2}{l^2}.
\end{align}
The coordinates range are
\begin{align*}
    -\infty<\,t&<\infty,\\
    0\leq \,r&<-\infty,\\
    0\leq \,\tilde{\theta}&\leq\pi/2,\\
    0\leq\,\tilde{\psi}&<4\pi,\\
    0\leq\,\tilde{\phi}&<2\pi.
\end{align*}
There is a singularity at $\rho=0$, and an event horizon occurs for $r=r_{+}$, the largest positive root of $\Delta_r=0$. When no such real root exists, the singularity is naked. The limiting case is the extremal solution, where all positive real roots of the equation coincide. The condition for it to happen is $\Delta_r(r_+)=0$ and $\partial_r\Delta_r(r)|_{r=r_+}=0$. 

Further, there are two $U(1)$ symmetries, associated with the two independent rotation angles, $(\tilde{\phi},\tilde{\psi})$, which generates two conserved angular momenta. The angular momenta are proportional to the rotation parameters $(a,b)$, ranging from $0\leq a,b<l$. Alongside, there is also a time translation symmetry.

Important quantities are the angular velocities of the spacetime points, $ (\Omega_\psi,\Omega_\phi)$, associated with the two rotation angles. They can be calculated at any point as being the angular velocity of a particle with zero angular momentum. Nonetheless, relevant for us are the velocities at the horizon, $(\Omega_\psi|_H,\Omega_\phi|_H)$ and at the conformal boundary $(\Omega_\psi|_{bdy},\Omega_\phi|_{bdy})$, which are found to be
\begin{align}
    \Omega_\phi|_H=&\frac{a-a^3/l^2}{a^2+r^{2}_{+}},&\hfill \Omega_\psi|_H=&\frac{b-b^3/l^2}{b^2+r^{2}_{+}}\\
    \Omega_\phi|_{bdy}=&-\frac{a}{l^2},& \hfill \Omega_\psi|_{bdy}=&-\frac{b}{l^2}.
\end{align}

Notice that, if $a,b\geq l$, the boundary would spin faster than light. The thermodynamic relevant angular velocities $\tilde{\Omega}$ (the ones contributing to the first law) are the horizon velocities relative to the conformal boundary, i.e., $\tilde{\Omega}=\Omega|_H-\Omega|_{bdy}$, given by
\begin{equation}
     \tilde{\Omega}_\phi=\frac{a(1+r_+^2/l^2)}{a^2+r^{2}_{+}},\qquad\tilde{\Omega}_\psi=\frac{b(1+r_+^2/l^2)}{b^2+r^{2}_{+}}.
\end{equation}

The metric in the conformal boundary is the leading order term in a large $r$ expansion, where the line element goes as $\rmd s^2=\textrm{e}^{2\Phi}\rmd s^2|_{bdy}+\mathcal{O}(r^0)$, with $\textrm{e}^{\Phi}$ the conformal factor at the boundary. In the present case:

\begin{equation}
    \textrm{e}^{2\Phi}\rmd s^2|_{bdy}=\frac{r^2}{l^2}\left[-\rmd t^2+l^2\left(\frac{\rmd\tilde{\theta}^2}{\Delta_{\tilde{\theta}}}+\frac{\mathrm{sin}^2\tilde{\theta}}{\Xi_a}\rmd\tilde{\phi}^2+\frac{\mathrm{cos}^2\tilde{\theta}}{\Xi_b}\rmd\tilde{\psi}^2\right)+2\rmd t\left(a\frac{\mathrm{sin}^2\tilde{\theta}}{\Xi_a}\rmd\tilde{\phi}+b\frac{\mathrm{cos}^2\tilde{\theta}}{\Xi_b}\rmd\tilde{\psi}\right)\right],
\end{equation}
where $\textrm{e}^{\Phi}=r/l$.

Therein it becomes evident that, for this choice of coordinates, the boundary is spinning and the constant time slices does not describe a $S^3$. Therefore, it will be useful in what follows to use a chart in which the boundary is conformal to a static spacetime with three-spheres as constant time hypersurfaces. With this purpose, we change coordinates as
\begin{equation}
    \mathrm{tan}^2\tilde{\theta}=\frac{\Xi_a}{\Xi_b}\tan^2\theta,\qquad\tilde{\phi}=\phi-\frac{a}{l^2}t,\qquad\tilde{\psi}=\psi-\frac{b}{l^2}t.
\end{equation}

The transformed line element reads
 \begin{multline}
     \rmd s^2=-\frac{1}{\Delta_{\theta}}\left(\frac{r^2}{l^2}+1\right)\rmd t^2+\frac{\rho^2}{\Delta_\theta}\rmd \theta^2+\frac{\rho^2}{\Delta_r}\rmd r^2+\frac{r^2+a^2}{\Delta_\theta}\mathrm{sin}^2(\theta)\rmd\phi^2+\frac{r^2+b^2}{\Delta_\theta}\mathrm{cos}^2(\theta)\rmd\psi^2\\+\frac{2M}{\rho^2\Delta^2_\theta}\left(\rmd t-a\,\mathrm{sin}^2(\theta)\rmd\phi-b\,\mathrm{cos}^2(\theta)\rmd\psi\right)^2,
     \label{MP_ds^2}
 \end{multline}
where now $\Delta_\theta=1-a^2\mathrm{sin}^2(\theta)/l^2-b^2\mathrm{cos}^2(\theta)/l^2$. Here, it is patent that the angle $\theta$ indicates the location of the rotation axes, being at $\theta=0$, and at $\theta=\pi/2$ for rotations along $\phi$, and $\psi$, respectively.

As desired, the boundary line element becomes just
\begin{equation}
    \textrm{e}^{2\Phi}\rmd s^2|_{bdy}=\frac{r^2}{l^2\Delta_{\theta}}\bigg[-\rmd t^2+l^2\Big(\rmd\theta^2+\mathrm{sin}^2(\theta)\rmd\phi^2+\mathrm{cos}^2(\theta)\rmd\psi^2\Big)\bigg].
    \label{staticbdy_ds2}
\end{equation}

The conformal boundary is static and the constant time surfaces are three-spheres in Hopf coordinates with radius $l$. On the other hand, the conformal factor $\textrm{e}^{\Phi}=r/(l\sqrt{\Delta_\theta})$ now has a dependence in the angular variable $\theta$.

That the boundary is static can also be seen by inspection of the angular velocities, which become now

\begin{align}
    \Omega_\phi|_H=&\frac{a(1+r_+^2/l^2)}{a^2+r^{2}_{+}},&\qquad \Omega_\psi|_H=&\frac{b(1+r_+^2/l^2)}{b^2+r^{2}_{+}}\\
    \Omega_\phi|_{bdy}=&0,& \qquad \Omega_\psi|_{bdy}=&0.
\end{align}

So that the thermodynamic relevant angular velocity now coincides with the horizon's one, $\tilde{\Omega}=\Omega|_H$.

\subsection{Thermodynamical properties}

 Important thermodynamic quantities are defined at the horizon, which is generated by the Killing vector field $\xi_H=\partial_{t}+\Omega_\psi|_H\partial_{\psi}+\Omega_\phi|_H\partial_{\phi}$. From it we calculate the surface gravity, and find the BH temperature to be
\begin{equation}
    T_{BH}=\beta^{-1}_{BH}=\frac{r^2_+\partial_r\Delta_r|_{r=r_+}}{4\pi(a^2+r^2_+)(b^2+r^2_+)}=\frac{r^4_+(a^2/l^2+b^2/l^2+1)-a^2b^2+2r^{6}_{+}/l^2}{2\pi \,r_+(a^2+r^2_+)(b^2+r^2_+)}.
\end{equation}

Notice that, the temperature vanishes for the extremal solution, where $\partial_r\Delta_r|_{r=r_+}=0$. Thence, in the parameter region where temperature would become negative, instead of a black hole, there is a naked singularity. Thereby, this solutions are excluded.

\begin{figure}[ht]
\begin{subfigure}[h]{0.55\linewidth}
\includegraphics[width=\linewidth]{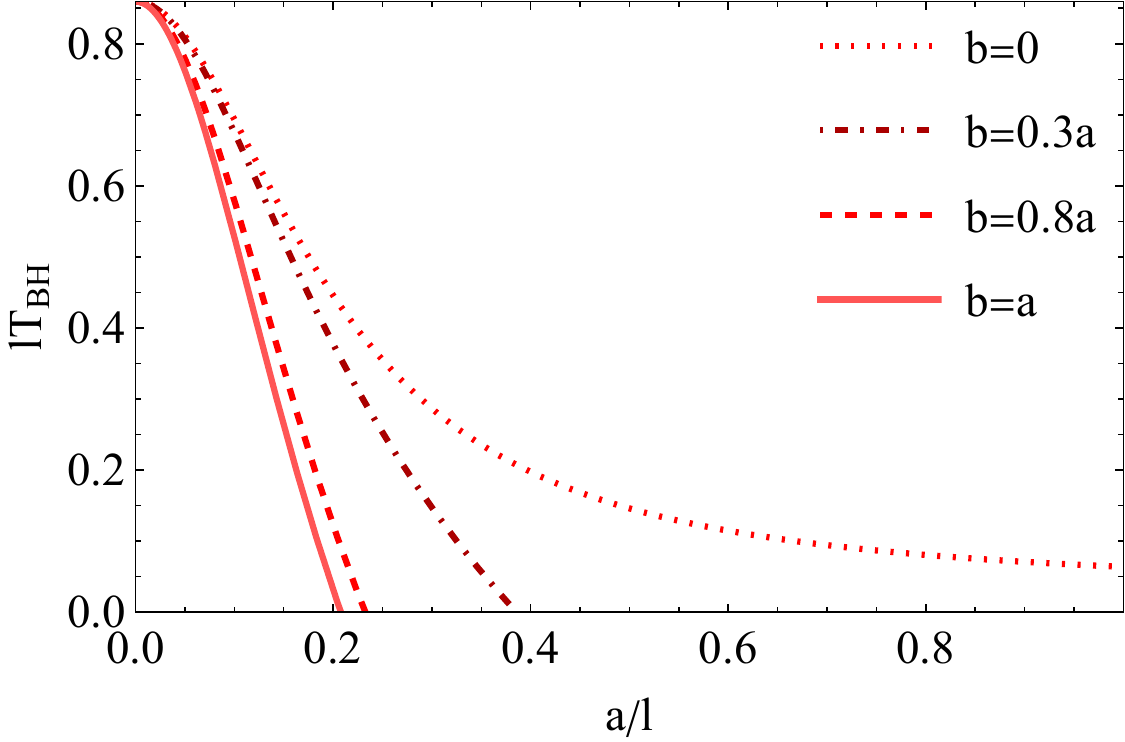}
\caption{$r_+=0.2l$.}
\end{subfigure}
\hfill
\begin{subfigure}[h]{0.55\linewidth}
\includegraphics[width=\linewidth]{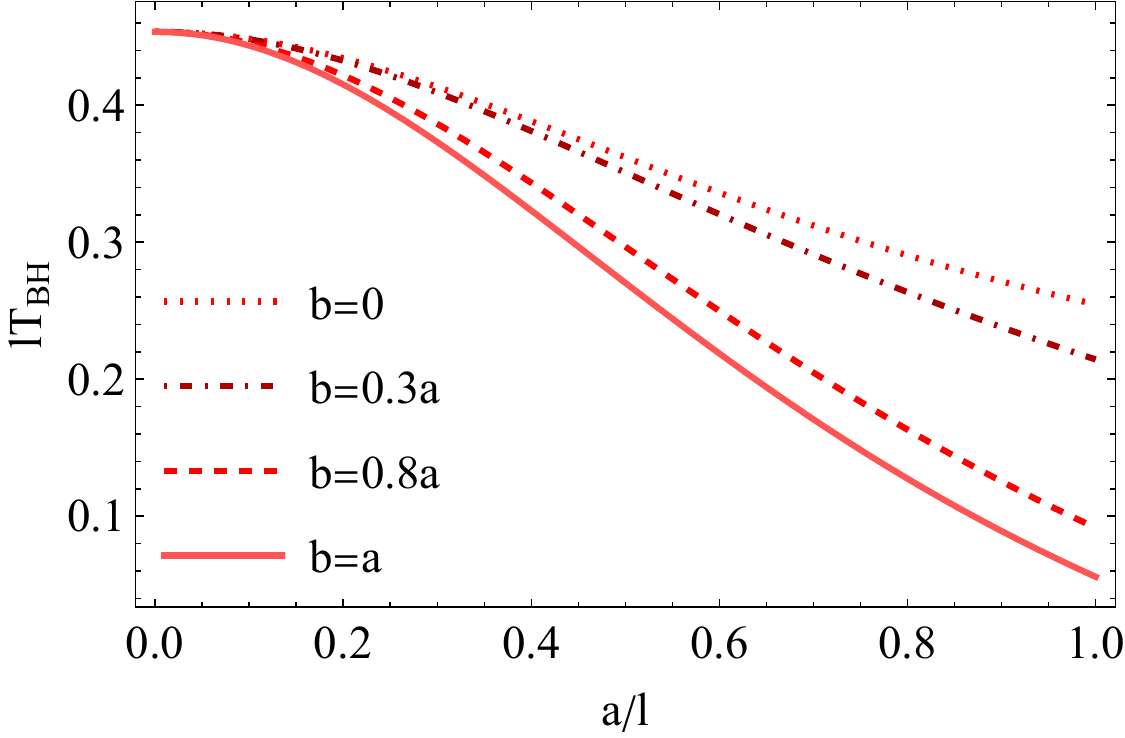}
\caption{$r_+=0.8l$.}
\end{subfigure}%
\newline
\begin{subfigure}[h]{0.55\linewidth}
\includegraphics[width=\linewidth]{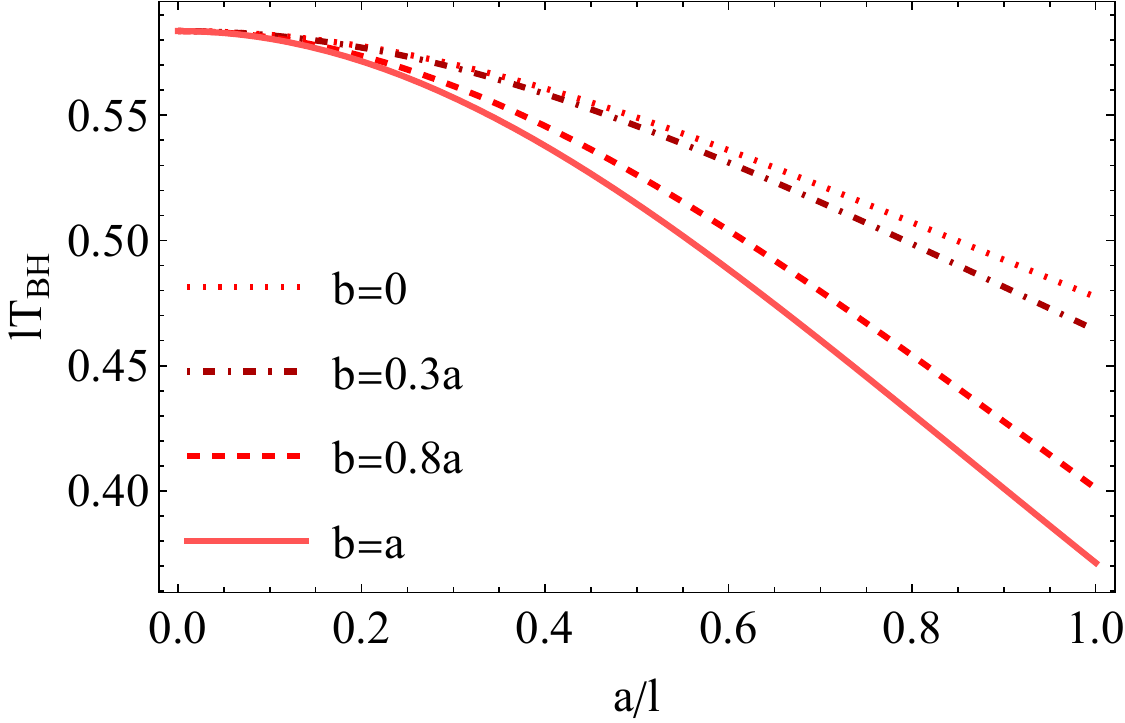}
\caption{$r_+=1.5l$.}
\end{subfigure}
\hfill
\begin{subfigure}[h]{0.55\linewidth}
\includegraphics[width=\linewidth]{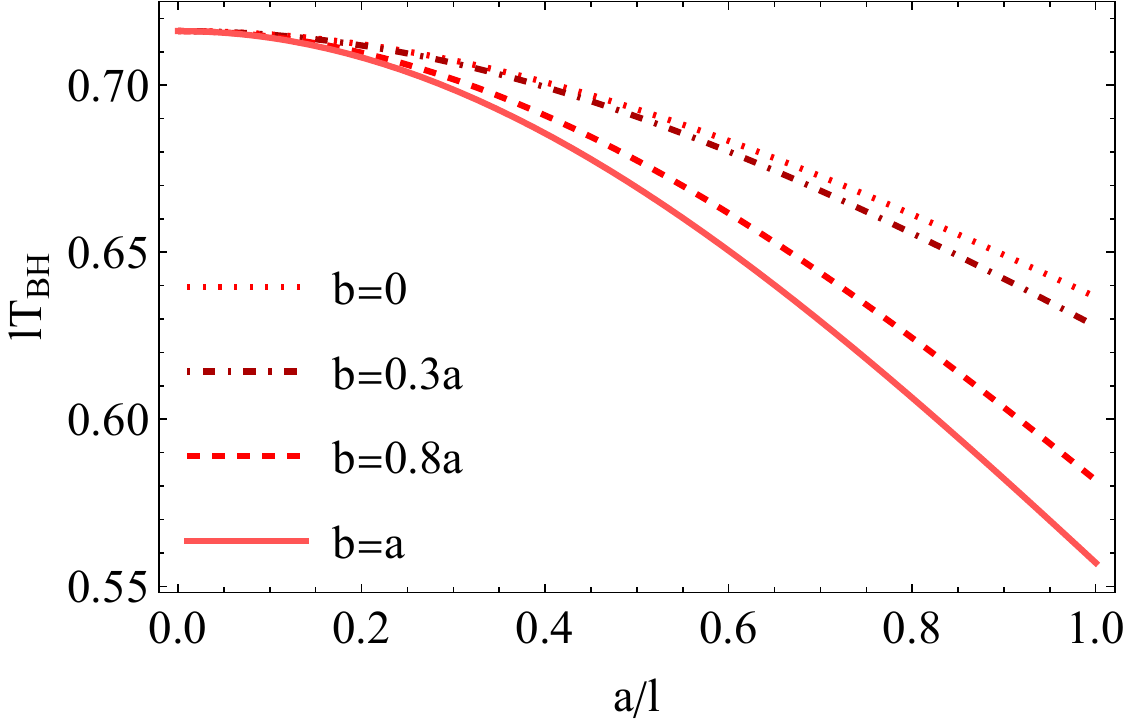}
\caption{$r_+=2l$.}
\end{subfigure}%
\caption{Plot of $lT_{BH}$ with respect to the rotation parameter $a/l$ for different choices of $(b/l,r_+/l)$. The temperature decays monotonically as the rotation parameter increases.}
\label{bhtempfig1}
\end{figure}

\begin{figure}[ht]
\begin{subfigure}[h]{0.55\linewidth}
\includegraphics[width=\linewidth]{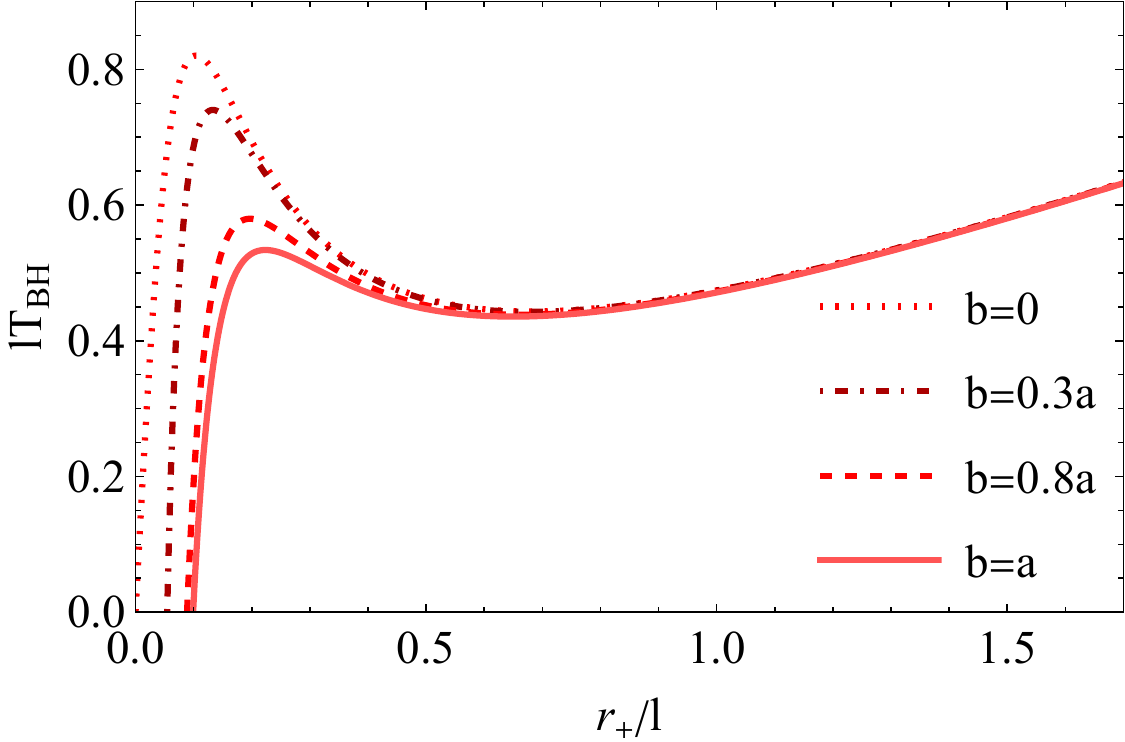}
\caption{$a=0.1l$.}
\end{subfigure}
\hfill
\begin{subfigure}[h]{0.55\linewidth}
\includegraphics[width=\linewidth]{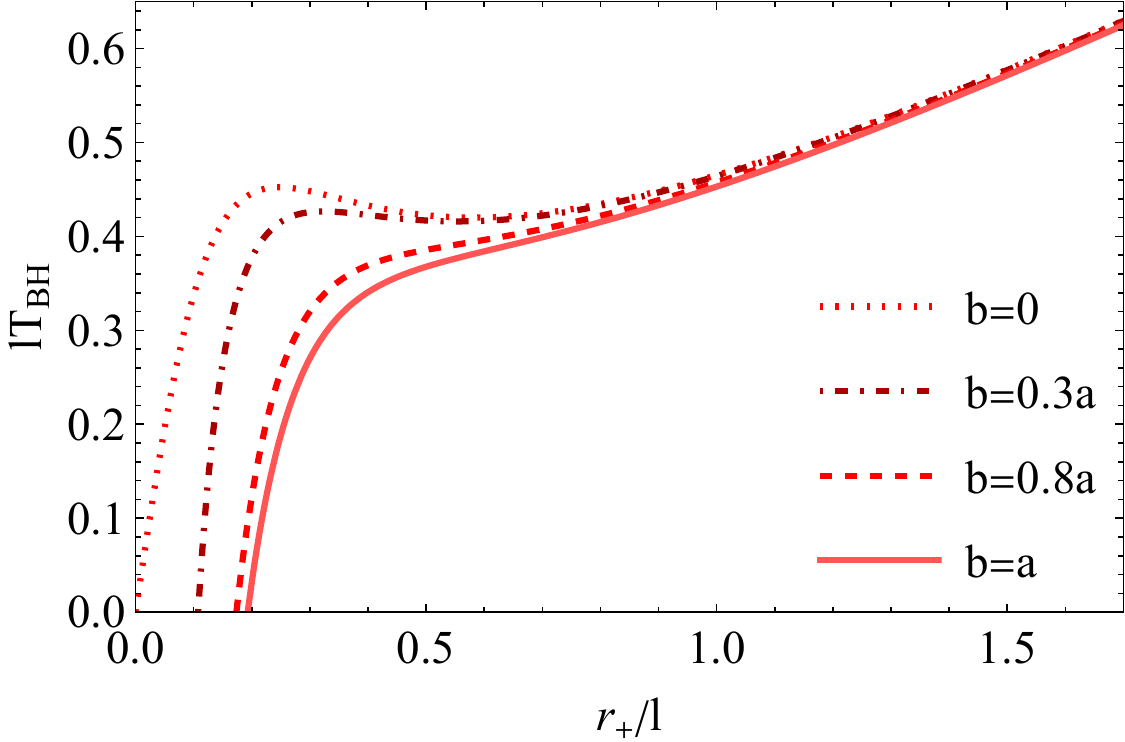}
\caption{$a=0.2l$.}
\end{subfigure}%
\newline
\begin{subfigure}[h]{0.55\linewidth}
\includegraphics[width=\linewidth]{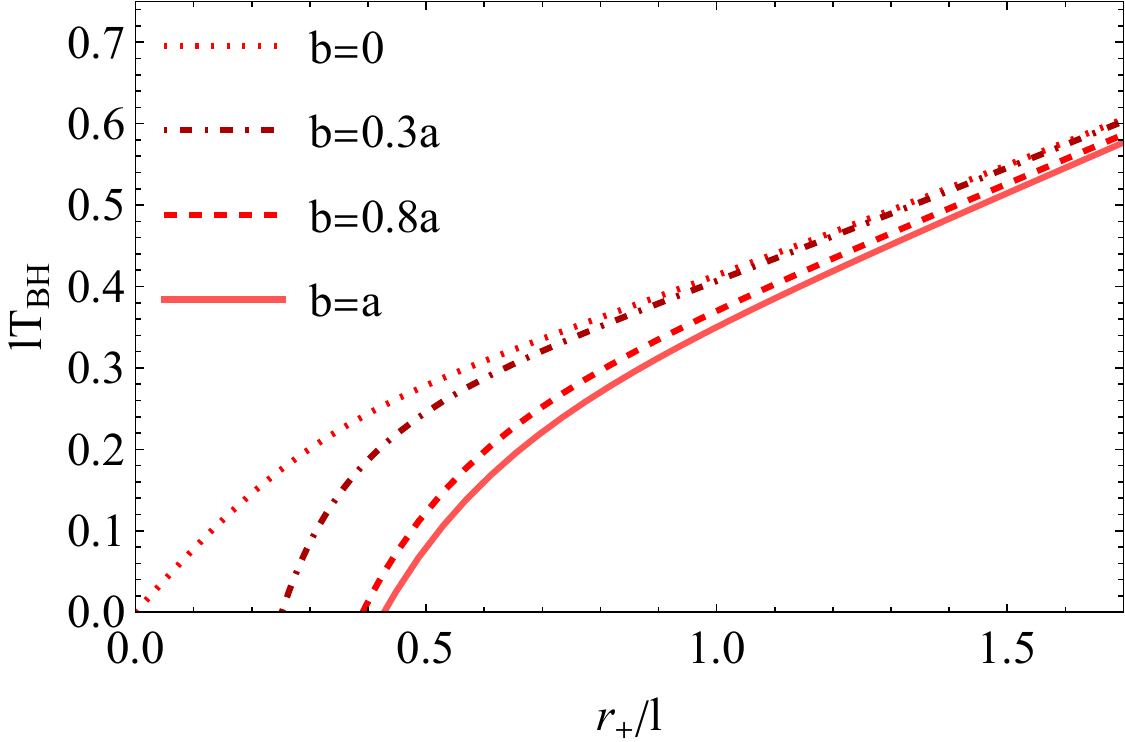}
\caption{$a=0.5l$.}
\end{subfigure}
\hfill
\begin{subfigure}[h]{0.55\linewidth}
\includegraphics[width=\linewidth]{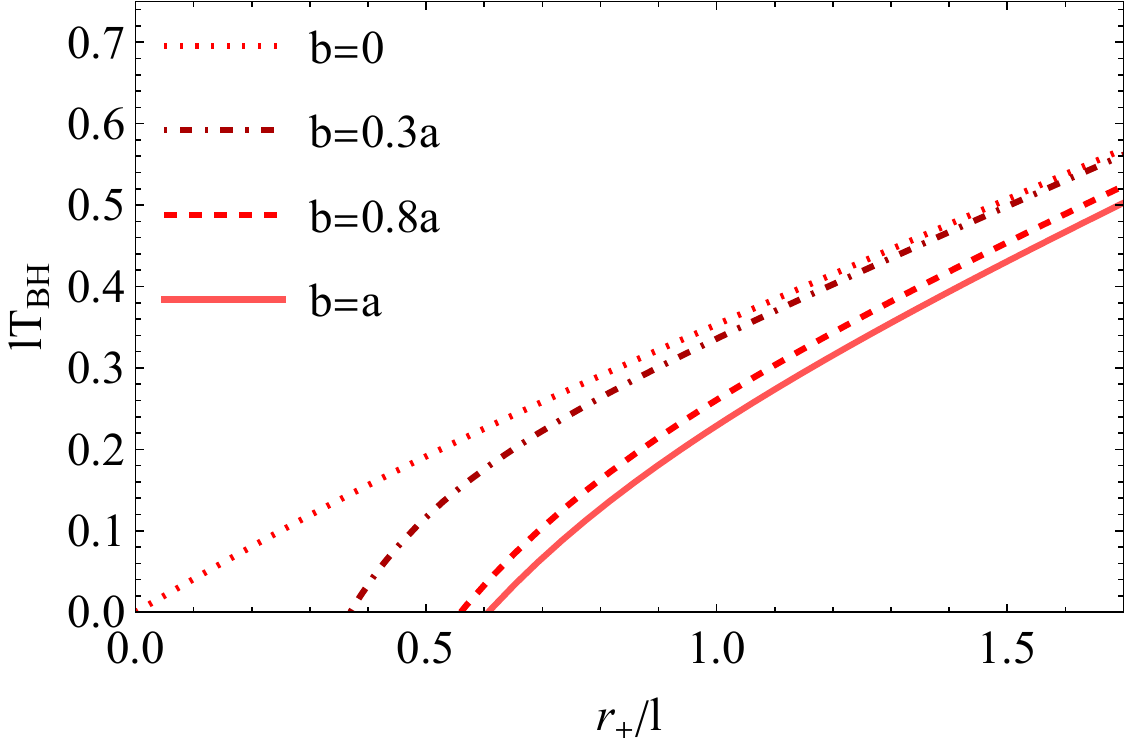}
\caption{$a=0.8l$.}
\end{subfigure}%
\caption{Plot of $lT_{BH}$ with respect to the horizon's radius $r_+/l$ for different choices of $(a/l,b/l)$. For small values of the rotation parameters, there is a non-monotonic profile of the temperature with increasing $r_+$. Otherwise, $T_{BH}$ increases with the horizon's radius.}
\label{bhtempfig2}
\end{figure}

The non-trivial temperature dependence on the parameters can be understood through the plots presented in Figure \ref{bhtempfig1}. There, it is shown how the temperature varies with the rotation parameter for different choices of $r_+$. In the absence of rotation, the temperature has a non-monotonic dependence in the horizon radius. As the rotation parameter increases the BH cools down monotonically, reaching the zero temperature solution only for small values of $r_+$. This behavior is reiterated in Fig. \ref{bhtempfig2}, where $lT_{BH}\times r_+/l$ is depicted for different choices of the rotation parameters. The temperature increases with the horizon, vanishing for a minimum value of $r_+$, that becomes larger as rotation increases.

Moreover, as in the absence of rotation, non-monotonic profiles of $T_{BH}$ are also present for small values of the rotation parameters. In such cases, two solutions with different values of $r_+$ will have the same temperature. This is associated with a first order phase transition between small and large black hole solutions \cite{Mann:2025xrb}. However, this phase transition is different from the Hawking-Page phase transition investigated here.

Finally, the entropy is proportional to the horizon's cross sectional area, and its given by
\begin{equation}
    S=\frac{4\pi^3(a^2+r^2_+)(b^2+r^2_+)}{\kappa^2l\,r_+\Xi_a\Xi_b},
\end{equation}
with $\kappa^2=8\pi G$. Additional extensive quantities of the spacetime are the angular momenta and the total energy, which can be calculated through the Komar formulas for the Killing vectors of the spacetime \cite{PS05} (For the energy a vacuum Casimir contribution must be subtracted). They are:
\begin{equation}
    J_\phi=\frac{4\pi^2 M}{\kappa^2\Xi^2_a\Xi_b}a, \qquad J_\psi=\frac{4\pi^2 M}{\kappa^2\Xi_a\Xi^2_b}b,
\end{equation}
\begin{equation}
    E=\frac{2\pi^2M(2\Xi_a+2\Xi_b-\Xi_a\Xi_b)}{\kappa^2\Xi_a^2\Xi_b^2}.
\end{equation}
These are seen to satisfy the first law of thermodynamics \cite{GPP05}:
\begin{equation}
    \rmd E=T\rmd S+\tilde{\Omega}_\psi\rmd J_\psi+\tilde{\Omega}_\phi\rmd J_\phi,
\end{equation}
where it appears the thermodynamic relevant angular velocity previously defined, $\tilde{\Omega}=\Omega|_H-\Omega|_{bdy}$. Different choices of angular velocity would not satisfy the first law \cite{GPP05}.

\subsection{Black brane and equal angular momentum cases}

It is of interest to work with a flat spacetime in the conformal boundary, instead of the $R\times S^3$ topology present in the full MP BH. To accomplish that, the usual procedure is to take the black brane limit of the black hole solution, i.e., perform a small angle, large radius approximation. Nonetheless, due to the broken symmetry, the expansion will depend on the choice of $\theta$, and a natural choice is to expand around one of the two axes of rotation. Take, for instance, the limit around $\theta=0$, accomplished through the coordinates $l(\phi,\theta)\to\epsilon(x,y)$, and $l\psi=z$. Additionally, one must rescale time and radius by $t\to\epsilon t$, $r/\Xi^{1/2}_a\to \epsilon^{-1}r$, and take the large horizon limit $r_+/\Xi^{1/2}_a\to \epsilon^{-1}r_+$. Then, for $\epsilon\to 0$, the line element becomes
\begin{equation}
    \rmd s^{2}=\frac{r^2}{L^2}\left[\frac{L^4\rmd r^2}{r^4-r^4_{+}}-\rmd t^2+\rmd x^2+\rmd y^2+\rmd z^2+\frac{r^4_+}{r^4(1-a^2/L^2)}\bigg(\rmd t+\frac{a}{L}\rmd x\bigg)^2\right],
    \label{bBB_ds2}
\end{equation}
being just the black brane solution boosted in the $x-$direction with parameter $a/l$ \cite{MM20}, the boundary now is a flat four dimensional spacetime. Despite that, any non-inertial effects are washed out, as rotation is reduced to a Lorentz boost. This means that the black brane approximation of the MP BH is not suited to investigate rotation. The described approximation could be made around the rotation axis in $\theta=\pi/2$, by taking $l(\psi,\theta-\pi/2)\to\epsilon(z,y)$, and $l\phi=x$, while rescaling the other quantities as before but with $\Xi_a\to\Xi_b$. Thence, the line element would be a black brane boosted in the $z-$direction with velocity $b/l$, similar to \eqref{bBB_ds2}.

Another widely used simplification is to take the equal angular momentum case of the MP BH, where $a=b$. There, we have that $\Delta_\theta=1-a^2/l^2=\Xi_a$, and $\rho^2=r^2+a^2$. The line element becomes simply
 \begin{multline}
     \rmd s^2=-\frac{1}{\Xi_a}\left(\frac{r^2}{l^2}+1\right)\rmd t^2+\frac{\rmd r^2}{\Delta_r|_{a=b}}+\frac{r^2+a^2}{\Xi_a}\left(\rmd\theta^2+\mathrm{sin}^2(\theta)\rmd\phi^2+\mathrm{cos}^2(\theta)\rmd\psi^2\right)\\+\frac{2M}{(r^2+a^2)\Xi_a^2}\left(\rmd t-a\,\mathrm{sin}^2(\theta)\rmd\phi-a\,\mathrm{cos}^2(\theta)\rmd\psi\right)^2.
     \label{equalMP_ds^2}
 \end{multline}
Therefore, the solution becomes spherically symmetric, invariant under the group $SU(2)\times U(1)$, larger than the two previous $U(1)$ symmetries. Both angular velocities are clearly the same, and one can have the solution spinning in only one direction by defining the new angles $\phi'=\phi-\psi$, and $\psi'=\phi+\psi$. The first, being the difference, does not spin, while the other has two times the previous angular velocity. The appearance of a spherical symmetry in the equal angular momentum case generates the situation mentioned in the introduction: all points in the constant radius three-spheres have the same linear velocity.

Here, we are interested in the broken symmetry scenario, with $a\neq b$, where a non-trivial $\theta-$dependence in the observables appears. Thereby, we shall work with the full MP BH solution described above. 

\section{Hawking-Page phase transition and confinement}
\label{sec3}

The AdS/CFT correspondence allows for a geometrical description of the confinement/deconfinement phase transition in strongly-coupled systems through the HP approach \cite{W98}. In this section we investigate the HP transition for a MP BH, calculating the confinement temperature. Then we discuss its dependence on the rotation parameters.    

\subsection{Hawking-Page transition in the MP BH solution}

The HP approach accounts for the phase transition between an asymptotic AdS black hole and a pure thermal AdS spacetime. For that, one assumes the existence of a gravitational partition function $\mathcal{Z}(\beta)$, with $\beta=T^{-1}$, the periodicity in compacted Euclidean time direction. Performing a semi-classical saddle point approximation, one has 
\begin{equation}
    \mathcal{Z}(\beta)=\int\mathcal{D}g\,\mathrm{e}^{-I_{E}[g]}\approx\mathrm{e}^{-I_{E}|_{on-shell}}.
\end{equation}
Here $I_{E}$ is the Euclidean gravitational action, being the Einstein-Hilbert action with a cosmological constant plus boundary terms:
\begin{equation}
    I_E=\frac{1}{2\kappa^2}\ \int^{\beta}_0\rmd\tau \int\rmd^{3}x\sqrt{g_E}\,(R-2\Lambda)+I_{GY},
\end{equation}
where a Wick rotation is performed in the time direction $t\to i\tau$, alongside with a redefinition of the rotation parameters $(a,b)\to i(\hat{a},\hat{b})$ for the Euclidean metric to remain real. All calculations performed here are at tree-level, being free from any analytic continuation problem. This is different from one or higher loop calculations, where poles in the real line forbid the continuation of imaginary angular velocities back to real values \cite{CFS22,CGM23,BCA24,C24}. Here, we will not write explicitly the Euclidean metric--it is straightforward to obtain it from the given line element--, and, therefore, from now on we will simply keep the real parameters $(a,b)$.  

Further, the on-shell action is a divergent quantity and we shall regularize it through the background subtraction method. In this regularization scheme, a pure thermal AdS on-shell action is subtracted from the BH one, canceling out any divergence. Moreover, as the BH contributions to the boundary term $I_{GY}$ decay rapidly, such term vanishes in the subtraction and we must only account for the contribution of the bulk action term. With that, the regularized free energy density of the BH will be given by
\begin{equation}
    \mathcal{E}_{BH}=\frac{1}{V_{bdy}}\lim_{\tilde{R}\to\infty}(I_{BH}-I_{AdS}),
\end{equation}
with $V_{bdy}=4\pi^2l^3$, the boundary spatial volume, and $\tilde{R}\to\infty$ an UV cutoff.

Now, we briefly describe the AdS spacetime, and a suitable choice of coordinates for it to be compared with the MP BH as presented last section. The usual line element for AdS space is
\begin{equation}
    \rmd s^2_{AdS}=-\left(\frac{y^2}{l^2}+1\right)\rmd t^2+\frac{\rmd y^2}{\frac{y^2}{l^2}+1}+y^2(\rmd \hat{\theta}^2+\mathrm{sin}^2\hat{\theta}\,\rmd\phi^2+\mathrm{cos}^2\hat{\theta}\,\rmd\psi^2).
    \label{AdS_ds^2}
\end{equation}
These coordinates are related to the Boyer-lindquist-like coordinates in \eqref{MP_ds^2} through the following transformation
\begin{align}
    y^2\mathrm{sin}^2\hat{\theta}&=(r^2+a^2)\mathrm{sin}^2\theta,\\ y^2\mathrm{cos}^2\hat{\theta}&=(r^2+b^2)\mathrm{cos}^2\theta.
\end{align}
In the new coordinates the line element \eqref{AdS_ds^2} becomes
\begin{equation}
    \rmd s_{AdS}^2=-\frac{1}{\Delta_{\theta}}\left(\frac{r^2}{l^2}+1\right)\rmd t^2+\frac{\rho^2}{\Delta_\theta}\rmd \theta^2+\frac{\rho^2}{\Delta_r|_{M=0}}\rmd r^2+\frac{r^2+a^2}{\Delta_\theta}\mathrm{sin}^2(\theta)\rmd\phi^2+\frac{r^2+b^2}{\Delta_\theta}\mathrm{cos}^2(\theta)\rmd\psi^2
    \label{AdS_ds^2_Bl}
\end{equation}
which is just the MP BH metric \eqref{MP_ds^2} with $M=0$.

Now we are able to compute the regularized free energy. The cosmological constant is $\Lambda=-
6/l^2$ and both have the same Ricci scalar, $R=20/l^2$, and Euclidean metric determinant
\begin{equation}
    \sqrt{g_E}=\frac{r\rho^2}{\Xi_a\Xi_b}\mathrm{sin}\theta\mathrm{cos}\theta.
\end{equation}
Thereby, the integral for the bulk action is
\begin{equation}
    I_{E}|_{on-shell}=\frac{16\pi^2}{\kappa^2 l^2\Xi_a\Xi_b}\int_0^\beta\rmd\tau\int ^{\tilde{R}}_{r_i}\rmd r\int^{\pi/2}_{0}\rmd\theta\, r\rho^2\mathrm{sin}\theta\mathrm{cos}\theta,
    \label{Eaction_int}
\end{equation}
The only difference between the spaces are the integration limits $(\beta,r_i)$. For the MP BH we have $\beta=\beta_{BH}$ and $r_i=r_+$, whilst for the pure AdS there are some subtleties. First, the geometries of both spaces must match in the asymptotic region $r=\tilde{R}\to\infty$ for the method to work. Therefore, the compactified time directions must have the same circumference, so that
\begin{equation}
    \sqrt{\Delta_r(\tilde{R})|_{M=0}}\beta_{AdS}=\sqrt{\Delta_r(\tilde{R})}\beta_{BH}\rightarrow\beta_{AdS}\approx\left(1-\frac{M}{\tilde{R}^4l^{-2}}\right)\beta_{BH}.
\end{equation}

Second, the lower limit of integration in the usual AdS coordinates \eqref{AdS_ds^2} is known to be the origin of the polar coordinates $y=0$, which is a coordinate singularity, but has a null measure when integrating over the plane $(y,\hat{\theta})$. On the other hand, in the new oblate spheroidal coordinates $(r,\theta)$, the point $y(r,\theta)=0$ becomes a surface and does not have a null measure anymore. Therefore, we introduce a trick to perform the integral in the plane $(r,\theta)$ in Eq. \eqref{Eaction_int}. For that, we go to cartesian coordinates $(x_1,x_2)$, such that
\begin{align}
    y^2\mathrm{sin}^2\hat{\theta}&=(r^2+a^2)\mathrm{sin}^2\theta=x_1^2,\\
    y^2\mathrm{cos}^2\hat{\theta}&=(r^2+b^2)\mathrm{cos}^2\theta=x_2^2.
\end{align}

Notice that, in the $(x_1,x_2)$ plane, a surface with constant $r$ is an ellipse. The lower limit of integration corresponds now to the origin $(x_1,x_2)=(0,0)$, and the upper is $r=\tilde{R}$, along all values of $0\leq\theta\leq\pi/2$. Thereby, the integration is over the area of an ellipse $C$ given by
\begin{equation}
    \frac{x^2_1}{\tilde{R}^2+a^2}+\frac{x^2_2}{\tilde{R}^2+b^2}=1.
\end{equation}
The integral becomes elementary:
\begin{equation}
    \int ^{\tilde{R}}_{r_i}\rmd r\int^{\pi/2}_{0}\rmd\theta\, r\rho^2\mathrm{sin}\theta\mathrm{cos}\theta=\int\int_{C}x_1x_2\rmd x_1\rmd x_2=\frac{1}{4}\int\int_{C}\rmd u\rmd t,
\end{equation}
being $u=x_1^2$, and $t=x_2^2$. Further, in the $(u,t)$ coordinates, the region under integration is just $C=\{(u,t)\in \mathbb{R} ^2:u>0,t>0,u+t(\tilde{R}^2+b^2)/(\tilde{R}^2+a^2)<1\}$, a triangle. Thence, the integral is simply the area of the triangle $C$:
\begin{equation}
    \frac{1}{4}\int\int_{C}\rmd u\rmd t=\frac{1}{8}(\tilde{R}^2+a^2)(\tilde{R}^2+b^2).
\end{equation}

Despite fairly simple, this trick can be straightforwardly generalized to higher dimension, giving an alternative computation method for the \textit{on-shell} action easier than the ones presetend previously (See appendix C in \cite{GPP05}).

Finally, with that we are able to compute the regularized free energy
\begin{multline}
     \mathcal{E}_{BH}=\frac{1}{\kappa^2l^3\Xi_a\Xi_b}\Big[M-l^{-2}(r^2_++a^2)(r^2_++b^2)\Big]=\frac{1}{\kappa^2l^5\Xi_a\Xi_b}(r^2_++a^2)(r^2_++b^2)(l^2-r^2_{+}).
\end{multline}
The most favorable thermodynamic system is the one with smaller free energy. As a consequence, $\mathcal{E}_{H}<0$ indicates that the black hole solution is stable, while for positive $\mathcal{E}_{H}$ the system is in a pure thermal AdS state. Therefore, the Hawking-Page phase transition between the BH and the AdS space occurs for $\mathcal{E}_{H}=0$, which implies the critical horizon radius $r^c_+=l$.

\subsection{Confinement/deconfinement critical temperature}

Through the AdS/CFT correspondence, we can identify the thermodynamical quantities of the black hole with the ones from the CFT at the boundary. In this context, Witten demonstrated that the BH corresponds to a high temperature deconfined phase of the dual SYM theory, while for the thermal AdS the dual system is confined \cite{W98}. Consequently, the HP approach offers a geometrical method to describe the confinement. Accordingly, the critical radius amounts to a critical temperature, given by
\begin{equation}
    \frac{T_c(a,b)}{T_c(0,0)}=\frac{4-(1-a^2/l^2)(1-b^2/l^2)}{3(1+a^2/l^2)(1+b^2/l^2)},
\end{equation}
with $T_c(0,0)=3/(2\pi l)$ the critical temperature in the absence of rotation. However, we want to investigate the profile of the critical temperature with respect to the angular velocities, not the rotation parameters. Then, notice that, at the critical horizon radius, the angular velocities become
\begin{equation}
    \tilde{\Omega}_\phi=\frac{2a/l^2}{(1+a^2/l^2)},\qquad\tilde{\Omega}_\psi=\frac{2b/l^2}{(1+b^2/l^2)}
\end{equation}
and the corresponding linear velocities are $(v_\phi,v_\psi)=(\tilde{\Omega}_\phi l,\tilde{\Omega}_\psi l)$. Substituting into the formula for the critical temperature we find that
\begin{equation}
    \frac{T_c(v_\phi,v_\psi)}{T_c(0,0)}=\frac{1}{3}\bigg(1+\frac{1}{\gamma_\phi}+\frac{1}{\gamma_\psi}\bigg),
\end{equation}
being $\gamma_\phi=(1-v_\phi^2)^{-1/2}$, and $\gamma_\psi=(1-v_\psi^2)^{-1/2}$, the Lorentz factor corresponding to each angular direction $(\phi,\psi)$.

\begin{figure}[!ht]
\includegraphics[width=0.8\textwidth]{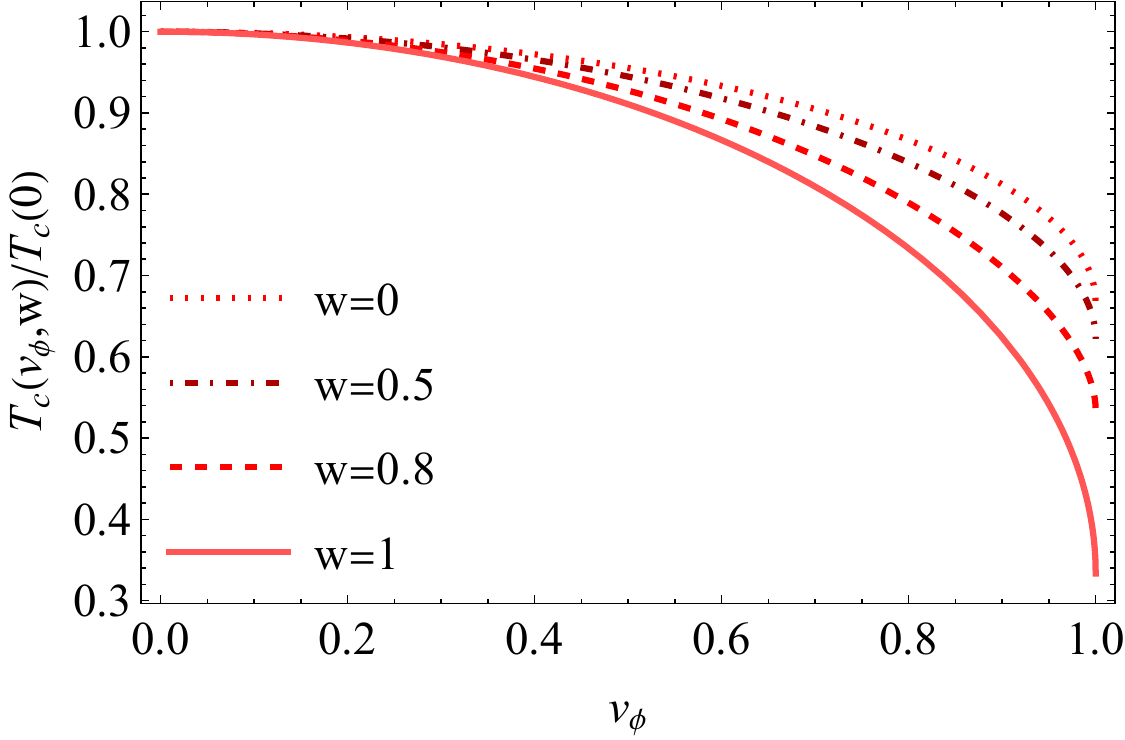}
\caption{Behavior of the critical temperature with respect to the linear velocity for different values of the ratio $w=v_{\psi}/v_{\phi}=\tilde{\Omega}_{\psi}/\tilde{\Omega}_{\phi}$. $T_c$ decays as the velocity increases, faster for greater values of $w$, with $w=1$ recovering the equal angular momentum case of Ref. \cite{Braga:2025wox}.}
\label{fig1}
\end{figure}
Thereby, rotation appears to ease deconfinement, as the critical temperature decreases with the velocity. Notice that the expression is symmetric under the exchange $v_\phi \leftrightarrow v_\psi$.

Further, it is interesting to introduce the dimensionless parameter
\begin{equation}
    w=\frac{v_\psi}{v_\phi}=\frac{\tilde{\Omega}_\psi}{\tilde{\Omega}_\phi}.
\end{equation}
Then, one can see that the critical temperature also decreases with $w$, decaying faster for the equal angular momentum case-- see Figure \ref{fig1}. This can also be seen by a small velocity expansion, which gives
\begin{equation}
    \frac{T_c(v_\phi,w)}{T_c(0,0)}=1-\frac{1}{6}(1+w^2)v_\phi^2+\mathcal{O}(v_\phi^4),
\end{equation}
therefore, greater the value of $w$, faster the decay, with $w=1$ recovering the equal angular momentum solution.

The temperature investigated in this section is the inverse of the $\beta$ parameter that goes into the partition function, and it cannot depend on any coordinate. This is what we call the reference temperature $T=\beta^{-1}$. Nonetheless, in the next section, we will demonstrate that this reference temperature coincides with the one measured by a static observer. This implies that the results presented above coincides with previous holographic models, and with Ref. \cite{Braga:2025wox}, where the critical temperature decreases with rotation when calculated in a static frame. In order to compare the model with lattice calculations, we will shall analyze the temperature as observed in a co-rotating frame, the same used in LQCD.

\section{Frame dependency of the local critical temperature}
\label{sec4}

One condition for thermal equilibrium is the absence of heat diffusion throughout the system. In general, this translates into a constant temperature distribution. Nonetheless, whenever gravitational fields or, equivalently, accelerations are present, the absence of diffusion requires a temperature gradient. Intuitively, one can think of heat transfer through radiation: Given a non-constant gravitational potential, photons would be red-shifted and lose energy, for instance, by climbing such potential; If all points had the same temperature, this would amount to an energy flux \cite{SV19}. This in turn implies that observers in different gravitational potentials will locally measure distinct temperatures.

This frame-dependency in the local temperature is drastic: in the equal angular momentum case, the confinement temperature has opposite behaviors in the static and in the co-rotating frames. In the former, it diminishes with rotation, while increasing in the latter \cite{Braga:2025wox}. Clearly, this is just a difference in the outcomes of temperature measurements for distinct observers, the state of the system is frame-independent: either a plasma or in a hadronic state. In this section, we investigate the local temperature in the strongly-coupled system dual to a general MP BH. Interestingly, there will appear a non-trivial $\theta-$dependence in the co-rotating temperature, due to the broken spherical symmetry. This is expected as the angle $\theta$ indicates the position of the rotation axes.

\subsection{Local temperature and confinement in a static frame}

The relation between the observer-dependent local temperature, $T^{loc}$, and the thermodynamic reference temperature $T$, is given by the Tolman-Ehrenfast law \cite{T30,TE30}:
\begin{equation}
    T^{loc}\sqrt{-g_{tt}|_{obs}}=T,
    \label{TElaw}
\end{equation}
where $g_{tt}|_{obs}$ is the component of the metric in the reference frame of the observer which performs the measurement.

As seen, the boundary metric is given by \eqref{staticbdy_ds2}, which was constructed to be static and, thereby, adapted to a non-rotating observer at the boundary. Nonetheless, there is a conformal factor that, in principle, should be taken into account in the formula \eqref{TElaw}, yielding $T^{loc}=\textrm{e}^{-\Phi}T_{BH}$. In the present case, as the conformal factor has a $\theta$-dependent term, which would imply the same dependence in the local temperature, found to be  $T^{loc}(\theta)=\sqrt{l\Delta_\theta/r }\,T_{BH}$.

However, in order to correctly describe the spacetime in which the dual fluid lives, one must get rid of the conformal factor by performing a conformal transformation at the boundary \cite{ACKW24}:
\begin{equation}
    g'_{ij}|_{(bdy)}\rmd x^i\rmd x^j=\mathrm{e^{-2\Phi}}g_{ij}|_{(bdy)}\rmd x^i\rmd x^j=-\rmd t^2+l^2\big(\rmd\theta^2+\mathrm{sin}^2\theta\rmd\phi^2+\mathrm{cos}^2\theta\rmd\psi^2\big).
\end{equation}

Under such transformation, the local temperature changes as $T^{loc}\to\mathrm{e}^{\Phi}T^{loc}=T_{BH}$ (See appendix A in \cite{Bhattacharyya:2007vs}). Therefore, the conformal transformation exactly cancels the conformal factor that would appear in the local temperature. In conclusion, the temperature as measured in a static frame coincides with the reference temperature. This is expected as in the static frame there is no non-inertial forces or, equivalently, no gravitational fields.  

\begin{figure}[t]
\begin{subfigure}[h]{0.55\linewidth}
\includegraphics[width=\linewidth]{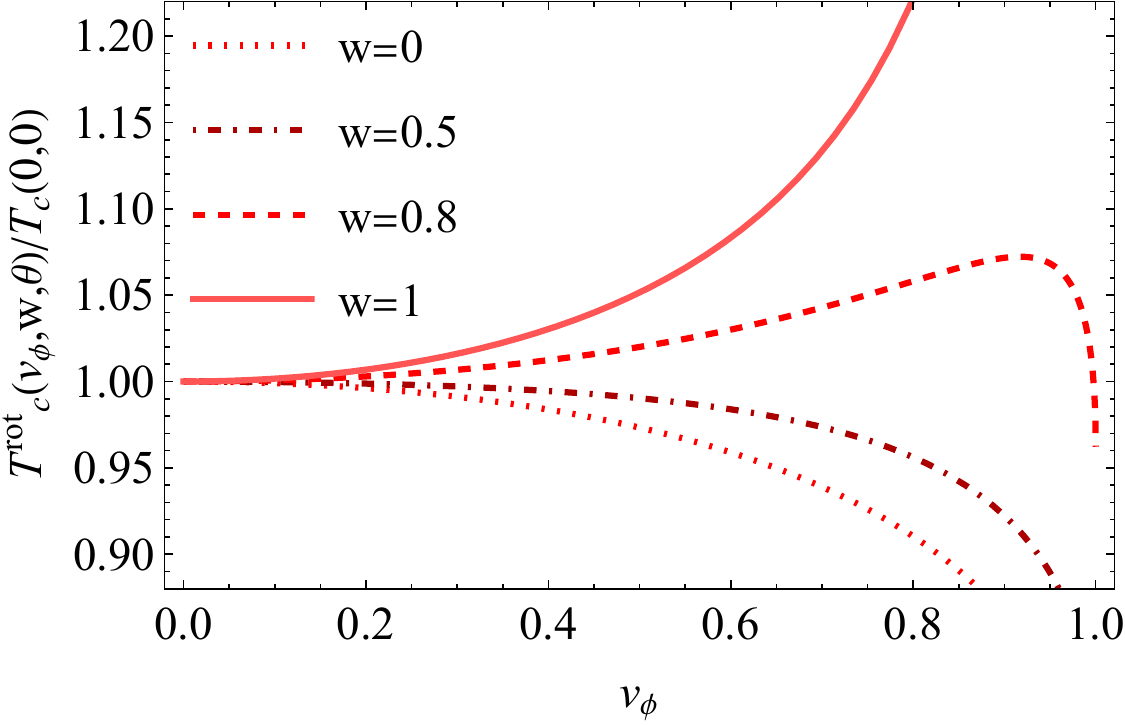}
\caption{$\theta=\pi/8$.}
\end{subfigure}
\hfill
\begin{subfigure}[h]{0.55\linewidth}
\includegraphics[width=\linewidth]{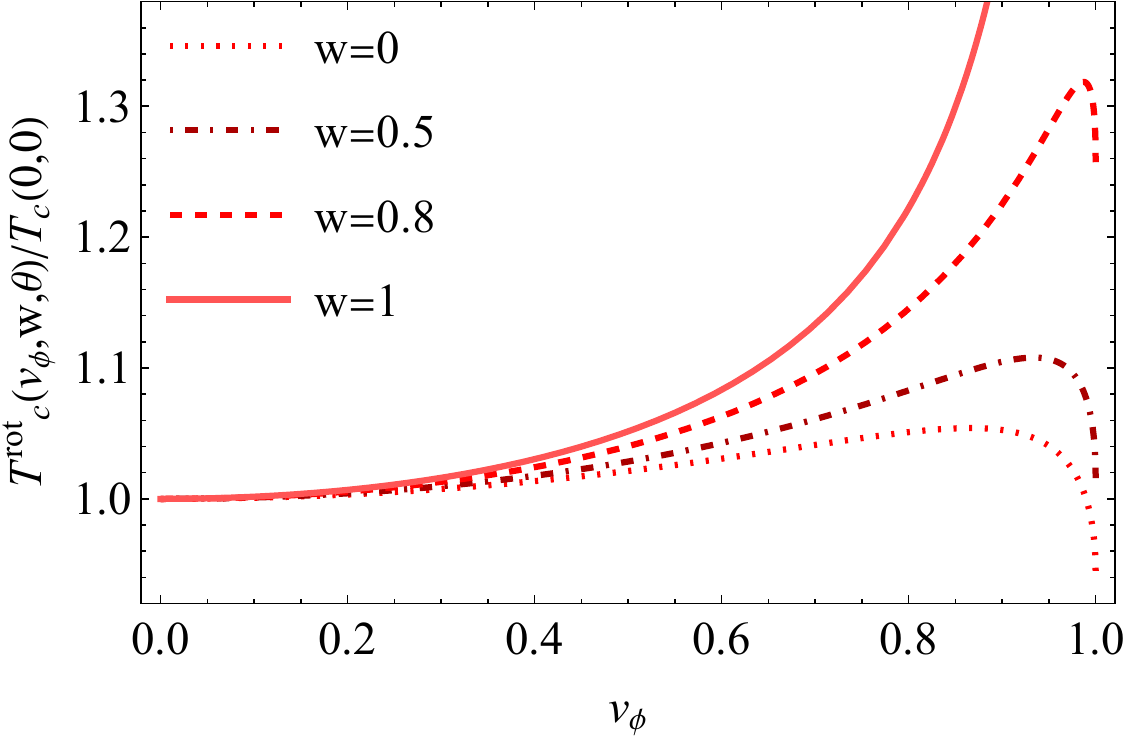}
\caption{$\theta=\pi/4$}
\end{subfigure}%
\label{Tloc1}
\caption{Behavior of the local critical temperature with respect to the velocity $v_\phi$ for different values of the ratio $w=v_\psi/v_\phi$, the angle $\theta$ is fixed at (a) $\theta=\pi/8$, and (b) $\theta=\pi/4$. The profiles change drastically for different values of $w$, in (a) they decrease for $w=0$ and $w=0.5$, being non-monotonic when $w=0.8$. However, for larger angles, as in (b), all curves increase initially. The only which increases monotonically is when $w=1$, which has no angular dependence.}
\label{fig4}
\end{figure}

\subsection{Local temperature and confinement in a co-rotating frame}

As discussed, LQCD calculations introduce rotation through the spacetime metric, by going to the co-rotating reference frame, where non-inertial effects are perceived as gravitational forces. This means that, in order to be able to compare our results with the lattice ones, both must be calculated in the co-rotating frame. For that, we must change the coordinates on the boundary to co-rotating ones through $(\phi,\psi)\to(\phi+\Omega_\phi t,\psi+\Omega_\psi t)$. The co-rotating boundary metric is
\begin{equation}
    \rmd s^2|_{bdy}=-\frac{\rmd t^2}{\gamma^2}+2l^2\rmd t\bigg(\Omega_\phi\mathrm{sin}^2(\theta)\rmd\phi+\Omega_\psi\mathrm{cos}^2(\theta)\rmd\psi\bigg)+l^2\bigg(\rmd\theta^2+\mathrm{sin}^2(\theta)\rmd\phi^2+\mathrm{cos}^2(\theta)\rmd\psi^2\bigg),
\end{equation}
where we define now the Lorentz factor of a co-rotating observer
\begin{equation}
    \gamma=\frac{1}{\sqrt{1-v_\phi^2\mathrm{sin}^2(\theta)-v_\psi^2\mathrm{cos}^2(\theta)}}.
\end{equation}
Now, the Tolman-Ehrenfast law gives the co-rotating temperature $T^{rot}=\gamma T_{BH}$, and its critical value is
\begin{equation}
\frac{T^{rot}_c(v_\phi,v_\psi,\theta)}{T_c(0,0)}=\frac{\gamma}{3}\left(1+\frac{1}{\gamma_{\phi}}+\frac{1}{\gamma_\psi}\right).
    \label{rot_temp}
\end{equation}
In the equal angular momenta case, $a=b$, the angular dependence vanishes and the results presented in \cite{Braga:2025wox} are recovered. In the broken symmetry scenario, where $a\neq b$, the $\theta$-dependence of the co-rotating temperature introduces interesting features not present in previous investigations. Notice also that, instead of the interchange symmetry between $v_\phi$ and $v_\psi$, the temperature is now symmetric under $(v_\phi,\theta)\leftrightarrow(v_\psi,\theta+\pi/2)$.

Some conclusion can directly be drawn by noting that $\gamma\geq1$, implies $T^{rot}_c\geq T_c$. The equality occurring only in the single-spinning case and in the axis of rotation. This is in accordance with the profile of the critical co-rotating temperature depicted in Figure \ref{fig4}. 
There, the co-rotating confinement temperature is shown as a function of the linear velocity for two different choices of angles. The equal angular momentum case shows an increase of the temperature, as obtained in \cite{Braga:2025wox}, while for different choices of the ratio $w$, the temperature may increase or not with rotation. Interestingly, when $T_c^{rot}$ starts increasing with $v_\phi$, it eventually decreases for near-luminal velocities, the only exception being the equal angular momentum case, $w=1$, or when $\theta=\pi/2$. This happens because, when $a\neq b$, $\gamma_\phi$, and $\gamma_\psi$ diverges faster than $\gamma$. Therefore, $(\gamma/\gamma_{\phi},\gamma/\gamma_{\psi})\to(0,0)$, when $(v_\phi,v_\psi)\to1$. 

\begin{figure}[t]
\begin{subfigure}[h]{0.55\linewidth}
\includegraphics[width=\linewidth]{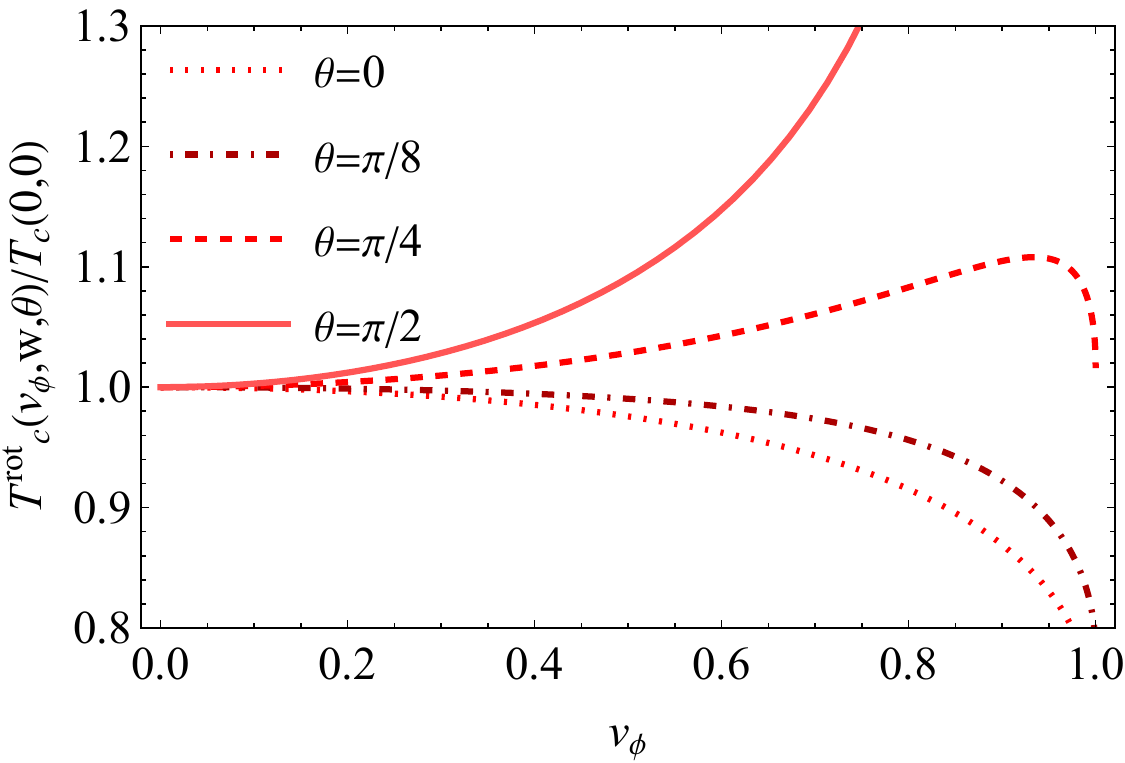}
\caption{$w=0.5$.}
\end{subfigure}
\hfill
\begin{subfigure}[h]{0.55\linewidth}
\includegraphics[width=\linewidth]{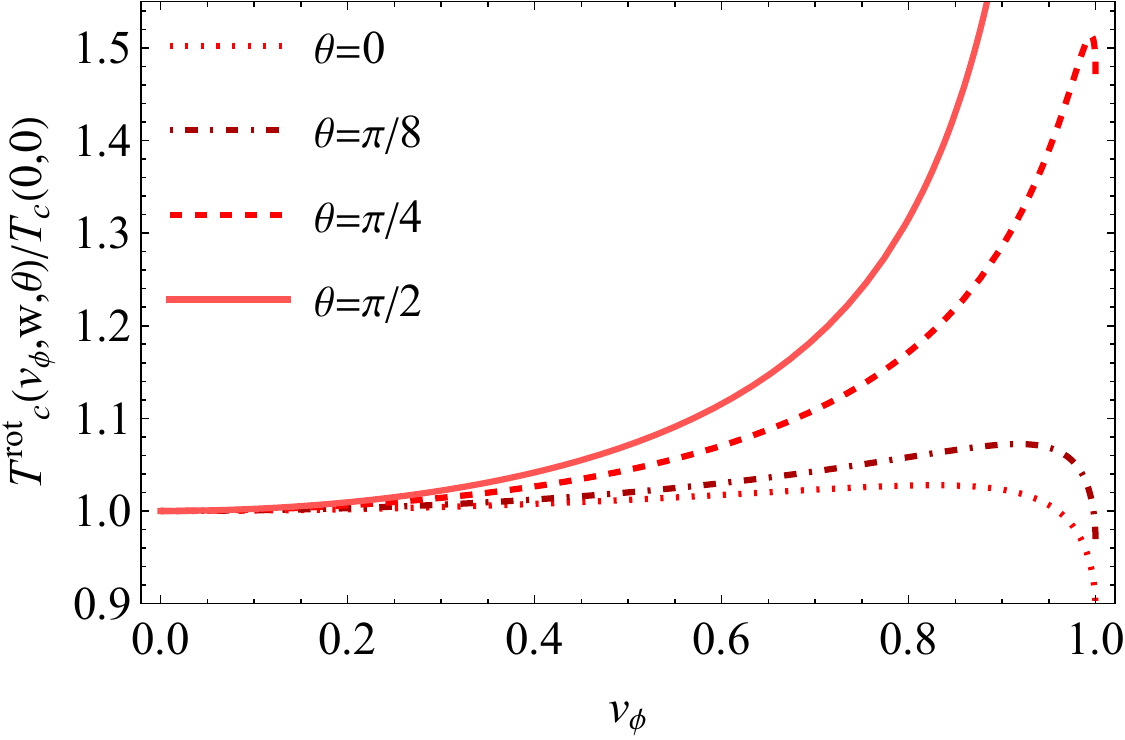}
\caption{$w=0.8$.}
\end{subfigure}%
\caption{Behavior of the local critical temperature with respect to the velocity $v_\phi$ for different values of the angle $\theta$, the ratio $w=v_\psi/v_\phi$ is fixed at (a) $w=0.5$, and (b) $w=0.8$. These plots corroborate the previous conclusion, for smaller angles and $w=0.5$, the critical temperature decreases. While for higher values of $w$, the temperature initially increases, decaying for near-luminal velocity. The only exception is for $\theta=\pi/2$, when it increases monotonically.}
\label{fig5}
\end{figure}

Some insight is gained by introducing again the parameter $w=v_\psi/v_\phi$, and performing a small velocity expansion, so that
\begin{equation}
    \frac{T^{rot}_c(v_\phi,w,\theta)}{T_c(0,0)}=1+B_2(w,\theta)v_\phi^2+\mathcal{O}(v_\phi^4),
    \label{Trot_smallv}
\end{equation}
where
\begin{equation}
    B_2(w,\theta)=\frac{1}{6}\Big[1-(1-w^2)(3\,\mathrm{cos}^2\theta-1)\Big].
\end{equation}

For the critical temperature to increase with velocity in this approximation, we must have that $B_2>0$, which implies that
\begin{equation}
    \mathrm{cos^2\theta}<f(w^2)=\frac{2-w^2}{3(1-w^2)}.
    \label{ineq_smallvel}
\end{equation}
Where the function $f(\omega^2)\in[2/3,\infty)$ for $w^2\in[0,1]$, increasing monotonically. Therefore, as $f(1/2)=1$, for values of $w>1/\sqrt{2}\approx0,707$, i.e $v_\psi>\sqrt{2}v_\phi$, the inequality \eqref{ineq_smallvel} is always satisfied and the critical temperature increases with a small rotation. However, when $w<1/\sqrt{2}$, values of $\theta$ smaller than a threshold value $\tilde{\theta}$, given by
\begin{equation}
    \tilde{\theta}=\mathrm{arccos}\left(\sqrt{\frac{2-w^2}{3(1-w^2)}}\right),
\end{equation}
give a decreasing critical temperature. The opposite holds for $\theta>\tilde{\theta}$.

To have some physical intuition take, for instance, the single-spinning case, i.e., $w=0$, then $\gamma=(1-v_\phi^2\mathrm{sin}^2(\theta))^{-1/2}$. Thereby, for small values of $\theta$, the Lorentz factor approaches unity and the co-rotating temperature behaves as the static one, meaning that the confinement temperature decreases with rotation.

\begin{figure}[t]
\begin{subfigure}[h]{0.55\linewidth}
\includegraphics[width=\linewidth]{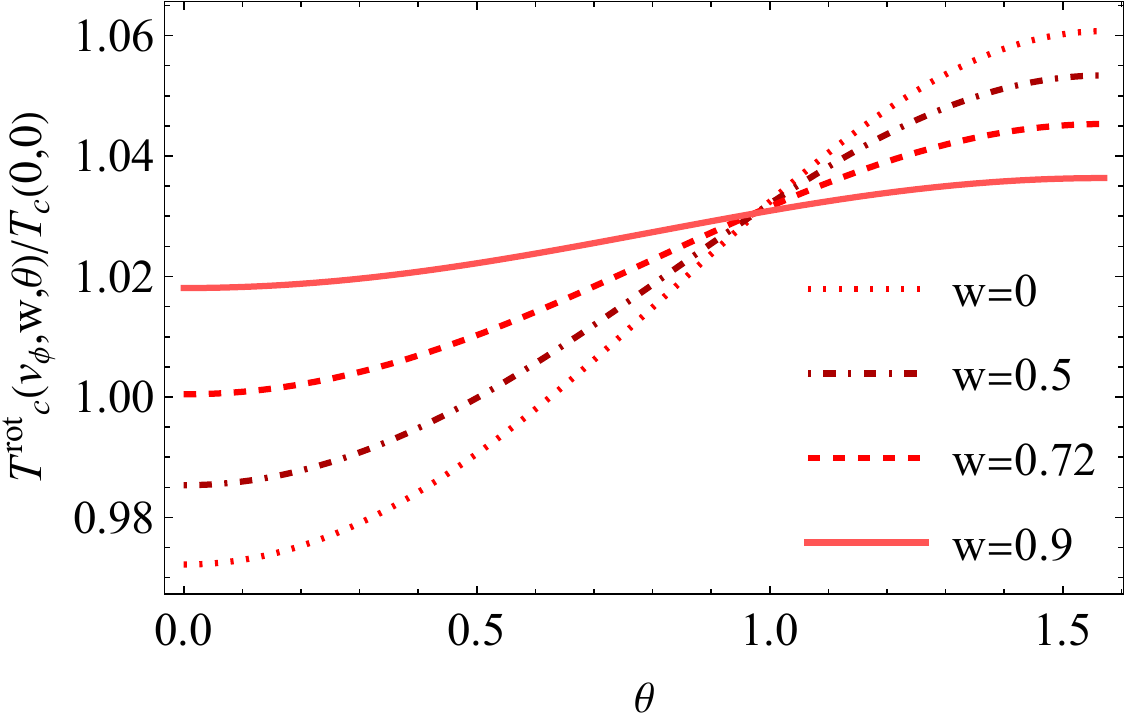}
\caption{$v_\phi=0.4$.}
\end{subfigure}
\hfill
\begin{subfigure}[h]{0.55\linewidth}
\includegraphics[width=\linewidth]{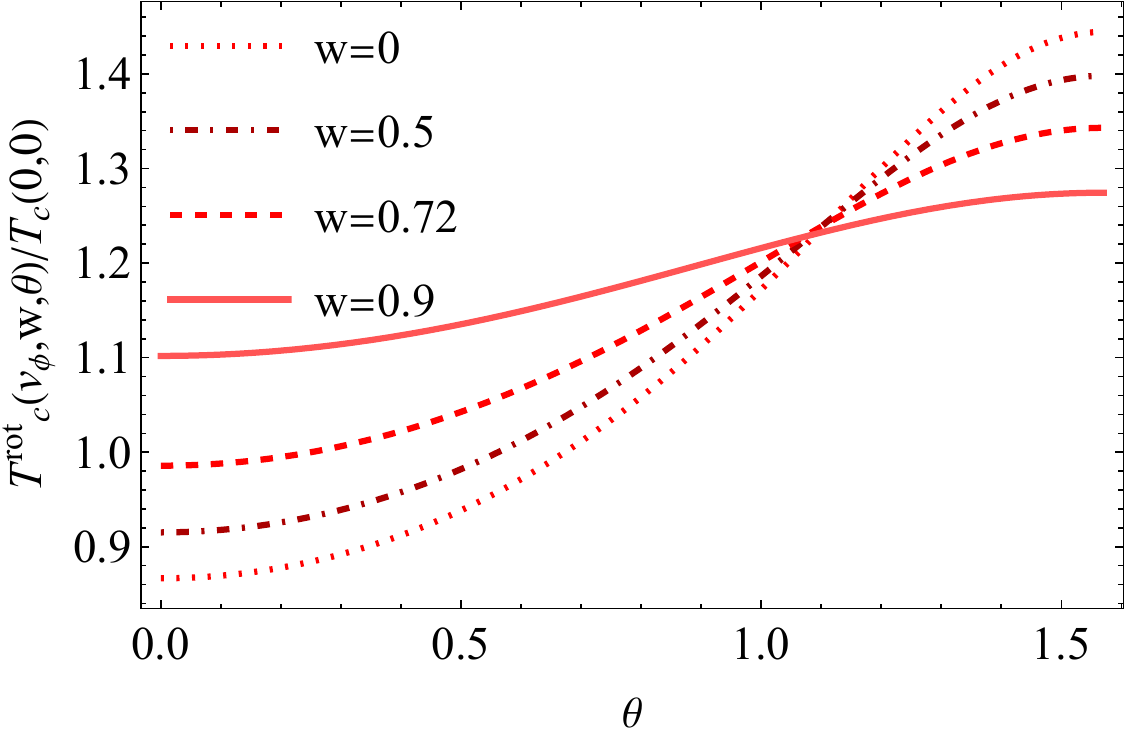}
\caption{$v_\phi=0.8$.}
\end{subfigure}%
\label{Tloc3}
\caption{Behavior of the local critical temperature with respect to the angle $\theta$ and the the ratio $w=v_\psi/v_\phi$, for fixed values the velocity, being (a) $v_\phi=0.4$, and (b) $v_\psi=0.8$. One can see that the $\theta-$dependence of the critical temperature is smaller for smaller velocities and greater values of $w$. In (a), the temperature starts smaller than unity for $w<0.72$, as expected. While in (b), it starts smaller even for $w=0.72$, as now the small velocity expansion is not valid.}
\label{fig6}
\end{figure}

Further, it is interesting to investigate the behavior for near-luminal velocities. For that we define $\varepsilon^2=1-v_\phi^2$, and expand the critical temperature in this new variable, to find
\begin{equation}
    \frac{T^{rot}_c(\varepsilon,w,\theta)}{T_c(0,0)}=\frac{1}{\sqrt{1-w^2\mathrm{cos}^2\theta-\mathrm{sin}^2\theta}}\left(1+\sqrt{1-w^2}+\varepsilon\right)+\mathcal{O}(\varepsilon^2).
\end{equation}

A divergence appears for $w=1$ or $\theta=\pi/2$, cases in which the expansion is not valid. Aside from these values, notice that $\partial T^{rot}_c/\partial\varepsilon>0$, up to first order. Therefore, since an increasing velocity implies a decreasing $\varepsilon$, the critical temperature decreases with rotation in the near-luminal regime. 

Further, when $\theta=\pi/2$ the near-luminal expansion becomes:
\begin{equation}
    \frac{T^{rot}_c(\varepsilon,w,\pi/2)}{T_c(0,0)}=\frac{1+\sqrt{1-w^2}}{3\varepsilon}+\frac{1}{3}+\mathcal{O}(\varepsilon),
\end{equation}
where the limit $w\to1$ can be taken. Moreover, as for $w=1$ the spherical symmetry is restored, this limit is independent of the angle $\theta$. In this case we have that $\partial T^{rot}_c/\partial\varepsilon<0$, and the temperature increases with $v_\phi$.

In Fig. \ref{fig5} we have $T^{rot}_c\times v_\phi$ for different angles, and one can see that the only case in which the temperature increases monotonically is for $\theta=\pi/2$. Moreover, when $w=0.5$, the critical temperature decreases for the smaller angles, while, for $w=0.8>1/\sqrt{2}$ it always increases in the small velocity regime, decreasing afterwards, in accordance with the previous discussion.

Finally, Fig. \ref{fig6} depicts $T^{rot}_c\times \theta$. The critical temperature always increases with the angle. When $v_\phi=0.4$ the departure from the non-rotating case is small and only for $w<1/\sqrt{2}$ the temperature for $\theta\approx0$ is smaller than one. However, for $v_\phi=0.8$, one can see that $w=0.72>1/\sqrt{2}$ is slightly smaller than unity for small angles, because then the small velocity approximation, and the related results, are not valid. Further, when $w$ increases, the change of the temperature with the angle is smaller, vanishing for $w=1$.

\section{Final Remarks}

In the present article, we analyzed the confinement/deconfinement transition in a rotating strongly-coupled system, dual to a MP BH solution. For that, we investigated the HP phase transition between the MP BH and a pure thermal AdS spacetime, which are dual to a plasma and a hadronic state, respectively. We then demonstrated that the local confinement temperature decreases with respect to the angular velocity, when measured by a static observer. Nonetheless, in the co-rotating frame, the qualitative behavior changes drastically for different values of the angle $\theta$: It can decrease, increase, or have non-monotonic profiles. The results are expected to have an angular dependence, as $\theta$ indicates the position of the rotation axes, located at $\theta=0$, and $\theta=\pi/2$, for rotations along the $\phi$, and $\psi$ directions, respectively.

The frame-dependency of the local temperature comes as no surprise. Rotation induces non-inertial forces that, akin to a gravitational field, induce a temperature gradient in thermodynamic equilibrium. As distinct observers may be subject to different inertial forces, they may measure different temperatures. The relation between the outcomes is given through the well-known Tolman-Ehrenfest law \cite{T30,TE30}, widely used in LQCD to investigate the spatial dependence of temperature \cite{BKKR21,CGM23,C24}. Conversely, we introduced it in the present work to discuss the frame-dependency of the temperature.

It is interesting to compare our work with previous results. In Ref. \cite{Braga:2025wox}, the equal angular momentum case was considered and it was found that the co-rotating critical temperature, $T^{rot}_c$ increases with rotation. This solution has spherical symmetry, and every point in the boundary spins with the same angular velocity, thereby measuring the same temperature. In the present case, instead, the symmetry is broken, and the co-rotating temperature depends on two novel parameters: the ratio $w=v_\psi/v_\phi=\tilde{\Omega}_\psi/\tilde{\Omega}_\phi$, and the angle $\theta$, indicating the position of the axes. Interestingly, the monotonically increasing confinement temperature is only recovered in the special cases when $\theta=\pi/2$, or when $w=1$, which corresponds to the equal angular momentum case. Besides that, the profile of $T^{rot}\times v_{\phi}$ is either monotonically decreasing or non-monotonic--increasing initially, but decreasing for near-luminal velocities. The temperature measured in a static frame is not significantly affected. It still decreases with angular velocity, faster for larger values of $w$, up to $w=1$.

In the co-rotating frame, our results can be confronted with LQCD. In the lattice, an expression for small velocities is found, as in eq. \eqref{Trot_smallv}, but with coefficient $B_2\approx0.7/1.3/0.5$, for open, periodic and Dirichlet boundary conditions, respectively \cite{BKKR21,BKKR22}. Our results are closer to LQCD for $w=1$ or $\theta=\pi/2$, as the expansion coefficient is larger, $B_2=1/6$, being the same found in Ref \cite{Braga:2025wox}. Nonetheless, as discussed, for $w>1/\sqrt{2}$, $B_2>0$ for any $\theta$ and the qualitative behavior of the lattice is reproduced.

Moreover, the broken spherical symmetry opens up the possibility to investigate, holographically, rotation induced inhomogeneous phases, as the ones that have been reported in the lattice \cite{BCA24}. Because, differently from the equal angular momentum case, points across the $S^3$ have distinct temperatures. Apparently, the model as presented here would not support inhomogeneities, as it is either a black hole or a pure AdS spacetime. Nonetheless, a construction similar to Ref. \cite{BJ24} could be done, where the plasma is given by a superposition of different gravitational duals, that can be in either one of the two states.

Further, when rotation is introduced through a boost, as in \cite{CZ21,BFJ22,Y23,WF24,ZH23}, only inertial effects are being considered, and the local co-rotating temperature is the same as in the zero angular momentum case. Besides that, angular velocity could also be introduced passively, through a rotation of the coordinate system, which can be different from the active case considered here \cite{Facundes:2025xcc}. Nonetheless, as discussed through the paper, the thermodynamic angular velocity is defined relatively to the boundary. If one rotates the coordinate system, the angular velocity of the boundary and of the horizon would be equally affected, and their difference would not change. Therefore, such a procedure seems to be innocuous. 

Summing up, we believe that our work proposes a protocol to investigate rotation using holography: first, angular momenta must be introduced actively, through the MP BH solution; second, the black brane limit cannot be taken, in order to account for non-inertial effects, which are essential; at last, the frame-dependency of the local temperature must be carefully taken into account. Remarkably, with this we can reproduce qualitatively the result found on the lattice in a given parameter region.
 
At last, a stereographic projection from $R\times S^3$ to $R^4$ is an approach alternative to the black brane limit in describing a plasma living on a flat space. In the case of a MP BH, where the resulting fluid is described by the Bantilan-Ishii-Romatschke flow \cite{Bantilan:2018vjv}, the projection not only retains the non-inertial effects, but also introduces several features expected to be present in the QGP produced at laboratory, such as vorticity and elliptical flow.

\acknowledgments

The authors are supported by FAPERJ -Fundaç\~ao Carlos Chagas Filho de
Amparo à Pesquisa do Estado do Rio de Janeiro. NRFB also acknowledges the support of  CNPq - Conselho Nacional de Desenvolvimento Científico
e Tecnológico and CAPES - Coordenação de Aperfeiçoamento de Pessoal de Nível Superior.




\bibliographystyle{JHEP}
\bibliography{biblio.bib}
\end{document}